\documentclass[pra,twocolumn,%showpacs,
amsmath,amssymb,nofootinbib]{revtex4}

\usepackage{epsfig}
\usepackage{color}
\usepackage{graphicx}
\usepackage{bm}
\usepackage{mathtools}
\usepackage{gensymb}
\usepackage{ulem}

\usepackage{fontenc}  % Prise en compte des accents

 \usepackage[colorlinks,bookmarks=false,citecolor=blue,linkcolor=red,urlcolor=blue]{hyperref}

\begin{document}

 \title{Why $\psi$ is incomplete indeed: a simple illustration}

\author{Philippe Grangier}

\affiliation{Laboratoire Charles Fabry, IOGS, CNRS, 
Universit\'e Paris Saclay, F91127 Palaiseau, France.}

\date{\today}

\date{\today}

\begin{abstract}
With the Nobel Prize attributed to Aspect, Clauser, and Zeilinger, the international scientific community acknowledged the fundamental importance of the experimental violation of Bell's inequalities \cite{Bell,Aspect}. It is however still debated what fails in Bell's hypotheses,  leading to these inequalities, and usually summarized as ``local realism'', or maybe more appropriately ``classical local realism''. The most common explanation is ``quantum non-locality", that remains however fully compatible with relativistic causality; this makes wondering whether any non-local phenomenon is really involved in these experiments. Here we want to recapitulate another option, sometimes called ``predictive incompleteness'', closely related to the idea that the usual state vector $\psi$ is incomplete indeed, as it was claimed by Einstein, Podolsky and Rosen \cite{ContextualInference}. However, the right way to complete $\psi$ has nothing to do with hidden variables, but requires to specify the measurement context, as it was claimed by Bohr \cite{debate}. Here we will consider the simple case of two spin 1/2, or two qubits, in order to keep the argument simple, but it does apply generally in quantum mechanics. 
\end{abstract}

\maketitle

Many discussions in Quantum Mechanics start with the statement ``let us consider the quantum state $|\psi \rangle$ of the system''. This allows one to initiate many calculations, but the question that should {\bf not} be asked by the beginner 
is ``but what is ``really" the quantum state $|\psi \rangle$ of the system''~? In practice, $|\psi \rangle$  is a vector in a Hilbert space, and it allows one to predict the future evolution of the system using a mathematical formalism found in QM textbooks, that has been vindicated in a huge number of experiments over more than a century \cite{CTDL}. 

For being concrete, and to remain in the context of Bell's inequalities, let us consider two spin 1/2 particles, so that $|\psi \rangle$ belongs to  a 4-dimensional Hilbert space, obtained as the tensor product of the 2-dimensional Hilbert spaces for each spin. As the formalism goes, one defines the spin operators $\vec S_1$ and  $\vec S_2$, with eigenvectors $|+ \rangle$ and  $|- \rangle$  for  the $S_z$ operators for each spin. A basis of the  2-spin Hilbert space is then usually denoted as $\{ |+ + \rangle, |+ -  \rangle, |- + \rangle, |- -   \rangle \}$ where the first $\pm$ in each ket refers to $\vec S_1$, and  the second one to $\vec S_2$.

Let us now consider a state in this space, the famous entangled singlet state $|\psi_s \rangle =  (|+ -  \rangle - |- + \rangle)/\sqrt{2}$. What does it mean to tell that ``the system is in state  $|\psi_s \rangle$'' ? In a naive way, is simply means that we can do some measurements on the pair of spin, that will give a result with certainty. This is actually true for any state in our Hilbert space, but it is particularly simple for the singlet state: it tells that if we measure the total angular momentum $\vec S$ of the two spins\footnote{The total spin $\vec S = \vec S_1 + \vec S_2$ is defined using the usual quantum rules for addition of angular momenta, and it has four orthogonal eigenstates denoted as $| S, M_S \rangle$, taking the values $\{ | 1,1 \rangle, | 1,0 \rangle, | 1,-1 \rangle, | 0,0   \rangle \}$, with  $|\psi_s \rangle =  | 0, 0 \rangle.$}, we will find 0 with certainty, for any component of $\vec S$ as well as for its modulus. 

This sounds like a fair definition of $|\psi_s \rangle$, but there is catch: if the system is in the state $|\psi_s \rangle$, we can measure many other quantities on the pair of spins, that will also give some results with certainty. For instance, one can perform a so-called Bell measurement, with the four orthogonal eigenstates $|\Phi^\pm \rangle  = (|+ + \rangle \pm |- -   \rangle)/\sqrt{2}$ and 
$ |\Psi^\pm \rangle = ( |+ -  \rangle \pm |- + \rangle)/\sqrt{2}$, and again $|\psi_s \rangle =   |\Psi^- \rangle$. There is actually an infinity of possible measurements, where $|\psi_s \rangle$ gives certain and reproducible results. Since all these measurements are generally incompatible (they are described by non-commuting operators), they clearly correspond to different physical situations. However,  these different  physical situations are not specified by giving $|\psi_s \rangle$, and therefore the only possible conclusion is that $|\psi_s \rangle$ is incomplete $\square$. 

In the usual formalism,  the specific operator corresponding to the quantity of interest has to be smuggled in somewhere. One sees thus how to complete $|\psi_s \rangle$: not by looking for any ``hidden variables'', but rather by specifying a particular measurement; or said otherwise, by specifying a basis of 4 orthonormal vectors including $|\psi_s \rangle$ among them.  We will call such a measurement a context, and the joint specification of some $|\psi \rangle$ and an associated context will be called a modality; said otherwise, a modality is the state of a system within a context~\cite{CO2002,csm1}. 

Taking the point of view that an actual physical state is a modality rather than a usual $|\psi \rangle$ is extremely helpful in practice \cite{psi}. It brings explicitly the idea that $|\psi \rangle$ is incomplete, and corresponds actually to an equivalence class of modalities belonging to different contexts \cite{extravalence,context}. It can also be said that specifying the context corresponds to the ``very conditions which define the possible types of predictions regarding the future behavior of the system'', as written by Bohr in his answer to Einstein, Podolsky and Rosen \cite{debate}. This brings the idea that $|\psi \rangle$ is ``predictively incomplete'':  this is an explicit way to violate Bell's inequalities, quite different from the possibility of non-local hidden variables \cite{ContextualInference}. Also,  the idea that $|\psi \rangle$ represents an equivalence class of modalities leads naturally to the hypotheses of Gleason's theorem, and thus to Born's rule \cite{csm4b,csm4c}. Finally, this point of view also leads to the possibility of reconsidering the formalism of operator algebra, less familiar than usual QM, but opening interesting possibilities to integrate both systems and contexts in a unified formalism \cite{completing,unification}. 

Overall, this point of view usually called CSM (Contexts, Systems and Modalities \cite{csm1})  is able, after setting up inductively some postulates based on the idea of contextual quantization \cite{CO2002,csm1}, to obtain deductively most of QM, including unitary transforms and Born's rule \cite{csm4b,csm4c}, and going up to a unified algebraic description of systems and contexts \cite{completing,unification}. 

From a more philosophical point of view \cite{debate}, these ideas rely on the notion of ``contextual objectivity" \cite{CO2002}, that is fully compatible with physical realism. In this view, a quantum  measurement does not require any conscious ``agent", but  does require a quantum system and a classical context. As an example, a Stern and Gerlach apparatus in a probe landed on a comet hundreds of millions km away is working just like it does on the earth. 
%\\

As a conclusion, the purpose of this short note is to underline that the violation of Bell's inequalities may point towards a view of QM that is more ``contextual" (and thereby non-classical) than ``non-local''. Though the context is a global concept it does not lead to any conflict with relativistic causality, because it allows for inferences, not influences \cite{notem}; this is explicitly shown for a Bell test by using a standard  light cone picture in \cite{ContextualInference}. 

This close relationship to contextuality \cite{context} leads to a complete reconsideration of the very notion of the properties of an isolated system: the classical notion of an isolated system owning well-defined properties is definitively lost in QM, and the objective physical object which carries such properties (now called modalities)  is a system within a context \cite{myst,def}. 
 \\

{\bf Acknowledgments} %======================================================================

The author thanks Alexia Auff\`eves, Nayla Farouki, and Mathias Van Den Bossche for useful comments.


\begin{thebibliography}{99}
\bibitem{psi} The present two-spin example can easily be extended to any  Hilbert space with a finite dimension larger than two. 
For continuous quantum variables the same considerations apply, provided that adequate care is taken when manipulating  e.g. the eigenstates of the position and momentum operators in a separable Hilbert space \cite{CTDL}.
\bibitem{notem} The apparent contradiction that the Universe should be considered as an isolated quantum system is solved by the fact that extremely large (and for modeling purposes mathematically infinite) quantum systems behave classically, thanks to von Neumann's sectorisation theorem, as explained in great details in \cite{unification}.
\end{thebibliography}
\end{document}